\documentclass[prb,twocolumn,superscriptaddress,showpacs]{revtex4-1}
\usepackage{graphicx, hyperref, multirow, verbatim, fontenc}

\begin{document}

\title{Layered structures of organic-inorganic hybrid halide perovskites}
\author{Tran Doan Huan}
\email{huan.tran@uconn.edu}
\affiliation{Institute of Engineering Physics, Hanoi University of Science and Technology, 1 Dai Co Viet Rd., Hanoi 100000, Vietnam}
\thanks{T.D.H. and V.N.T. contributed equally to this work.}
\affiliation{Department of Materials Science \& Engineering and Institute of Materials Science, University of Connecticut, 97 North Eagleville Rd., Unit 3136, Storrs, CT 06269-3136, USA}
\author{Vu Ngoc Tuoc}
\email{tuoc.vungoc@hust.edu.vn}
\affiliation{Institute of Engineering Physics, Hanoi University of Science and Technology, 1 Dai Co Viet Rd., Hanoi 100000, Vietnam}
\author{Nguyen Viet Minh}
\affiliation{Institute of Engineering Physics, Hanoi University of Science and Technology, 1 Dai Co Viet Rd., Hanoi 100000, Vietnam}

\begin{abstract}
Organic-inorganic hybrid halide perovskites, in which the A cations of an ABX$_3$ perovskite are replaced by organic cations, may be used for photovoltaic and solar thermoelectric applications. In this contribution, we systematically study three lead-free hybrid perovskites, i.e., methylammonium tin iodide CH$_3$NH$_3$SnI$_3$, ammonium tin iodide NH$_4$SnI$_3$, and formamidnium tin iodide HC(NH$_2$)$_2$SnI$_3$, by first-principles calculations. We find that in addition to the commonly known motif in which the corner-shared SnI$_6$ octahedra form a three-dimensional network, these materials may also favor a two-dimensional (layered) motif formed by alternating layers of the SnI$_6$ octahedra and the organic cations. These two motifs are nearly equal in free energy and are separated by low barriers. These layered structures features many flat electronic bands near the band edges, making their electronic structures significantly different from those of the structural phases composed of three-dimension networks of SnI$_6$ octahedra. Furthermore, because the electronic structures of HC(NH$_2$)$_2$SnI$_3$ are found to be rather similar to those of CH$_3$NH$_3$SnI$_3$, formamidnium tin iodide may also be promising for the applications of methylammonium tin iodide.
\end{abstract}


\maketitle

\section{Introduction}
Ternary compounds with the ABX$_3$ cubic perovskite structure, or simply perovskites, host an enormous number of functionalities, e.g., ferroelectricity, colossal magnetoresistance, and high thermopower. Interest in this family has further been fuelled by the development of organic-inorganic hybrid prerovskites in which A is replaced by an organic cation like methylammonium CH$_3$NH$_3$ or formamidinium HC(NH$_2$)$_2$. \cite{Burschka:2013, Gao:2014, Liu:perovskites, Kojima:perovskites, Hao2014, Noel:14, C001929A, SMLL:SMLL201402767, Lee_Science, Stoumpos_perovskite, perovskite_20pc, ADMA:ADMA201401137, Mitzi, Mitzi95, Xu:layer, Kagan99, Xu:03,Umebayashi:2003} While their photovoltaic applications with an efficiency of $\gtrsim 20\%$ \cite{perovskite_20pc} capture most of the recent attention,\cite{Burschka:2013, Gao:2014, SMLL:SMLL201402767, Liu:perovskites, Kojima:perovskites, Lee_Science, Stoumpos_perovskite, perovskite_20pc, Hao2014, Noel:14, C001929A, ADMA:ADMA201401137} some hybrid halide perovskites, e.g., CH$_3$NH$_3$PbI$_3$ and CH$_3$NH$_3$SnI$_3$, were also suggested to have large Seebeck coefficient. \cite{Stoumpos_perovskite, Takahashi201339} This implies that they may also be applicable for solar thermoelectric generators,\cite{Galli_perovskite} converting the concentrated sunlight into electricity using the Seebeck effect.\cite{Kraemer:Nat, Baranowski} These compelling applications promote a great deal of studies on this new class of materials, many of them are computation-based.\cite{Mosconi:persp:2015, Mosconi, Colella:2013, Galli_perovskite, Brivio:13, Brivio:perovskite, Frost:perovskite,Frost:nanolett,Umari_SR, KhuongOng,Mosconi:JMCA:2015}

Physical properties and hence, the efficiency of hybrid perovskites in such the applications are sensitive with their material structure, which turns out to be fairly complicated, as will be shown in this work. In the simple cubic ABX$_3$ perovskite structure, A cations sit at the centers of the cages formed by a three-dimensional (3D) network of connected BX$_6$ octahedra. When large, anisotropic and polar organic cations are introduced at these sites, this structure is dramatically deformed, resulting two new major structural motifs. The first motif, which is characterized by strongly distorted 3D network of BX$_6$ octahedra, has widely been examined by computations. \cite{Mosconi:persp:2015, Mosconi, Colella:2013, Galli_perovskite, Brivio:13, Brivio:perovskite, Frost:perovskite,Frost:nanolett,Umari_SR, KhuongOng,Mosconi:JMCA:2015} In the second motif, the 3D network is completely broken along one dimension, forming alternating layers of connected BX$_6$ octahedra and the organic cations. Consequently, structures of this two-dimensional (2D) layered motif are significantly different from the 3D structures.

Layered structures in hybrid perovskites are ubiquitous,\cite{Umebayashi:2003, Mitzi, Mitzi95, Xu:layer, Kagan99,Xu:03, C001929A} but computational effort devoted to them is surprisingly limited. In a work based on first-principles calculations,\cite{Umebayashi:2003} the experimentally resolved structure of C$_4$H$_9$NH$_3$PbI$_4$ was found to have a slightly wide band gap $E_{\rm g}$ accompanied with flat electronic band dispersion nearby. These characteristics were also suggested\cite{Mosconi} for several nearly-2D structures of CH$_3$NH$_3$PbBr$_3$ and CH$_3$NH$_3$PbCl$_3$, signaled by the calculated long/short/short/long pattern of Pb-X axial distances (X = Br, Cl). Different from the 2D structure examined for C$_4$H$_9$NH$_3$PbI$_4$,\cite{Umebayashi:2003} the 2D structures predicted \cite{Mosconi} for CH$_3$NH$_3$PbBr$_3$ and CH$_3$NH$_3$PbCl$_3$ share almost all the essential features of the 3D motif, except slightly longer Pb-X axial distances. In a series of hybrid perovskites with rather large organic cations, e.g., [NH$_2$C(I=NH$_2$)]$_2$(CH$_3$NH$_3$)$_m$ Sn$_m$I$_{3m+2}$,\cite{Mitzi95} [(CH$_3$)$_3$NCH$_2$CH$_2$NH$_3$]SnI$_4$, \cite{Xu:layer} (C$_6$H$_5$C$_2$H$_4$NH$_3$)$_2$SnI$_4$,\cite{Kagan99} and [CH$_3$(CH$_2$)$_{11}$NH$_3$]SnI$_3$, \cite{Xu:03} the layered motif is unambiguously dominant. Presumably, the 3D motif can not be preserved when such the large cations are introduced into the lattice. Even if the X anions are substituted by polyanions like BF$_4^-$ or PF$_4^-$, layered structures have also been suggested and studied computationally, revealing many flat electronic bands near the band edges.\cite{C4TA05284F}

In this paper, we focus on three organic-inorganic hybrid lead-free halide perovskites, namely methylammonium tin iodide CH$_3$NH$_3$SnI$_3$, ammonium tin iodide NH$_4$SnI$_3$, and formamidnium tin iodide HC(NH$_2$)$_2$SnI$_3$ at the level of density functional theory (DFT).\cite{DFT1,DFT2} By employing the minima-hopping structure prediction method, \cite{Goedecker:MHM,Amsler:MHM,MHM:OrganovBookChapter} we predict many low-energy structures for these hybrid perovskites, each of them exhibits ether the 2D or the 3D motif. For each material, the lowest-energy 2D and 3D structures are separated by a low energy barrier, roughly $30-40$ meV per atom. On the other hand, the 2D structures are characterized by flat electronic bands near the band edges, clearly differentiating themselves from the conventional 3D structures. This observation suggests that the layered structures of hybrid perovskites deserve further comprehensive investigations. We also find that the electronic structures of HC(NH$_2$)$_2$SnI$_3$ are rather similar to those of CH$_3$NH$_3$SnI$_3$, suggesting formamidnium tin iodide may also be applicable for the photovoltaic and thermoelectric applicatons of methylammonium tin iodide.

\begin{figure}[t]
  \begin{center}
    \includegraphics[width= 7 cm]{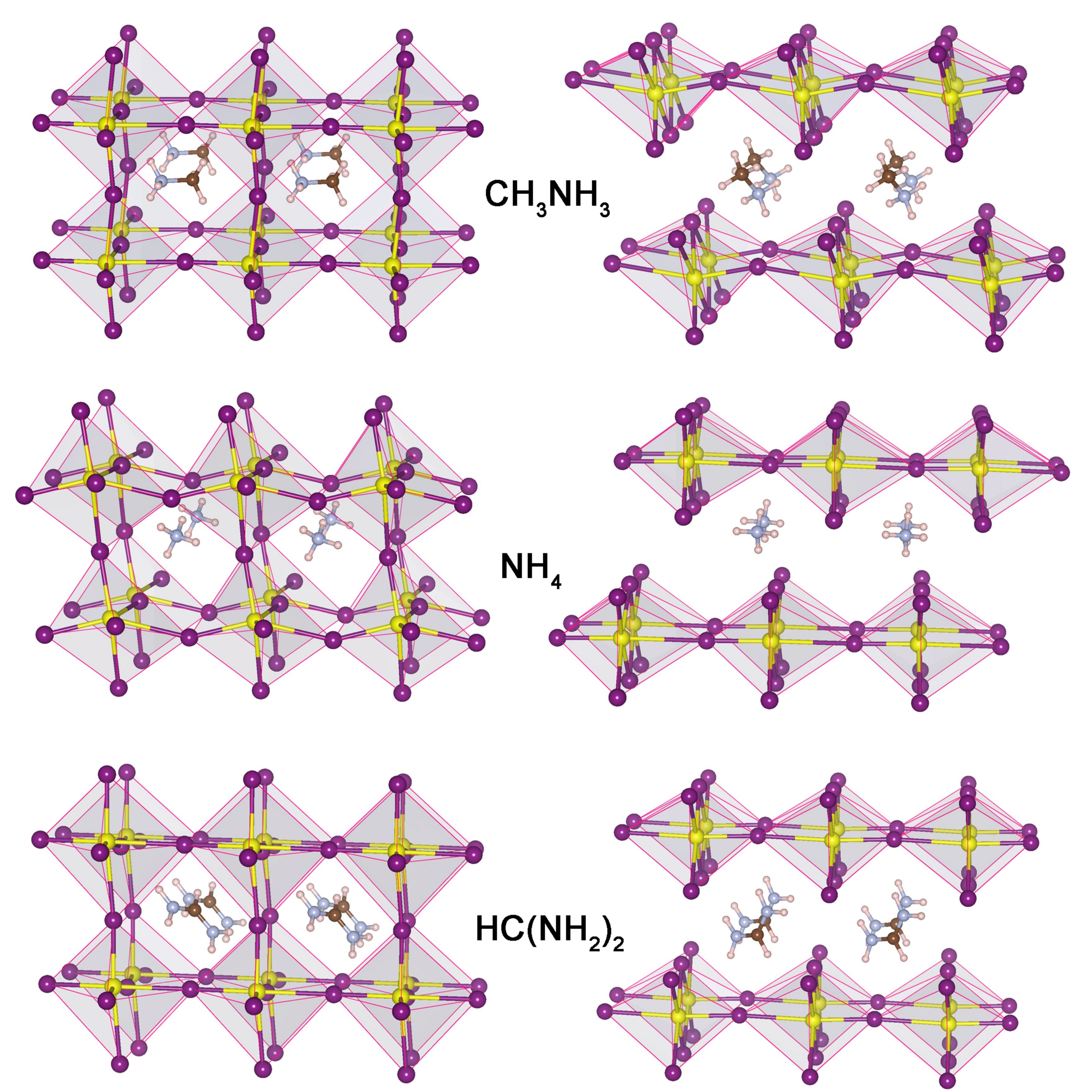}
  \caption{(Color online) Lowest-energy structures of CH$_3$NH$_3$SnI$_3$, NH$_4$SnI$_3$, and HC(NH$_2$)$_2$SnI$_3$: the 3D structure on the left and the 2D structure on the right. Sn, I, C, N, and H atoms are shown in yellow, purple, brown, lavender, and pink colors, respectively.} \label{fig:struct}
  \end{center}
\end{figure}

\section{Computational methods}
Our DFT calculations were performed with {\sc vasp} package, \cite{vasp1,vasp2,vasp3,vasp4} employing the projector augmented wave (PAW) formalism. \cite{PAW} A basis set of plane waves with kinetic energies up to 500 eV was used to represent the wavefunctions while the local density approximation (LDA) functional was used for the exchange-correlation (XC) energies. The possible relativistic effects from the spin-orbit coupling (SOC) were included in our calculations. The Brillouin zone of the examined structures was sampled by Monkhorst-Pack\cite{monkhorst} $\bf k$-point meshes of $7 \times 7 \times 7$. The equilibration of the examined structures was assumed when the residual atomic forces exerting on the ions are smaller than $10^{-2}$ eV/\AA.

Low-energy structures of the hybrid perovskites were predicted using the minima-hopping method. \cite{Goedecker:MHM,Amsler:MHM,MHM:OrganovBookChapter} Typically, a configurational space search is started by locally optimizing an initial structure by a DFT code ({\sc vasp} in this work). From the optimized structure, the energy landscape is then constructed from the DFT energy $E_{\rm DFT}$ and explored by multiple attempts, each of them consists of a molecular-dynamics run for escaping the current local minimum and a local geometry optimization run on the obtained structure. The molecular-dynamic trajectory is chosen along the low-curvature directions of the energy landscape for improving the prediction efficiency, as suggested by Bell-Evans-Polanyi principle. This is one of several feedback mechanisms and techniques that are employed in the minima-hopping method. This method has been successfully used for materials of various classes, including inorganic, \cite{Huan:Zn,Huan:Mixed,Huan:NaSc, Huan:hafnia} organic,\cite{Arun:design} and inorganic-organic hybrid compounds. \cite{Baldwin:SnEster1,Baldwin:SnEster2,Huan:Data}

\section{Low-energy structures}\label{sec:structs}
Many low-energy structures of CH$_3$NH$_3$SnI$_3$, NH$_4$SnI$_3$, and HC(NH$_2$)$_2$SnI$_3$ were predicted, exhibiting either the 3D motif or the 2D motif. In this work, we focus on six lowest-energy structures, one 2D and one 3D structure for each of the three materials. In Fig. \ref{fig:struct}, these six structures are visualized. Obviously, the topology of the ideal cubic perovskite structures is preserved in the structures of 3D motif, keeping them in a pseudo-cubic geometry, as experimentally described.\cite{Stoumpos_perovskite} On the other hand, layers of BX$_6$ octahedra are unambiguously formed in the 2D structures and strongly shifted with respect to each other. Moreover, the simulated x-ray diffraction (XRD) shown in Fig. 1 of the Supplemental Material,\cite{supplement} indicates that the examined 3D structures of CH$_3$NH$_3$SnI$_3$ and HC(NH$_2$)$_2$SnI$_3$ are essentially identical with the pseudo-cubic $\alpha$ phases experimentally resolved\cite{Stoumpos_perovskite} for these perovskites and computationally studied elsewhere.\cite{Galli_perovskite} In Sec. \ref{sec:elec}, this conclusion will further be confirmed by the excellent consistence between the calculated and the measured band gaps of CH$_3$NH$_3$SnI$_3$ and HC(NH$_2$)$_2$SnI$_3$. The crystallographic information and simulated XRD patterns of all the 2D and 3D structures can also be found in the Supplemental Material.\cite{supplement} The crystallographic information is also reported in \verb=http://khazana.uconn.edu=.

\begin{table}[t]
\caption{Calculated DFT energies $E_{\rm DFT}$ of the 2D structures of three hybrid perovskites, given in meV/atom, with respect to those of the 3D structures. Results were obtained by using various numerical prescriptions described in the text.}\label{table:energy}
\begin{center}
\begin{tabular}{l c c c c c}
\hline
\hline
Material              & LDA+SOC  & LDA     & PBE      & HSE06   & vdW-DF2  \\
\hline
CH$_3$NH$_3$SnI$_3$   &  $7$   &  $6$  &  $10$    & $7$   & $2$   \\
NH$_4$SnI$_3$         & $-6$   & $-7$  &  $0$     & $-2$  & $-15$ \\
HC(NH$_2$)$_2$SnI$_3$ &  $1$   & $-1$  &  $1$     & $-1$  & $-4$  \\
\hline
\hline
\end{tabular}
\end{center}
\end{table}

\begin{figure}[t]
  \begin{center}
    \includegraphics[width= 7.5 cm]{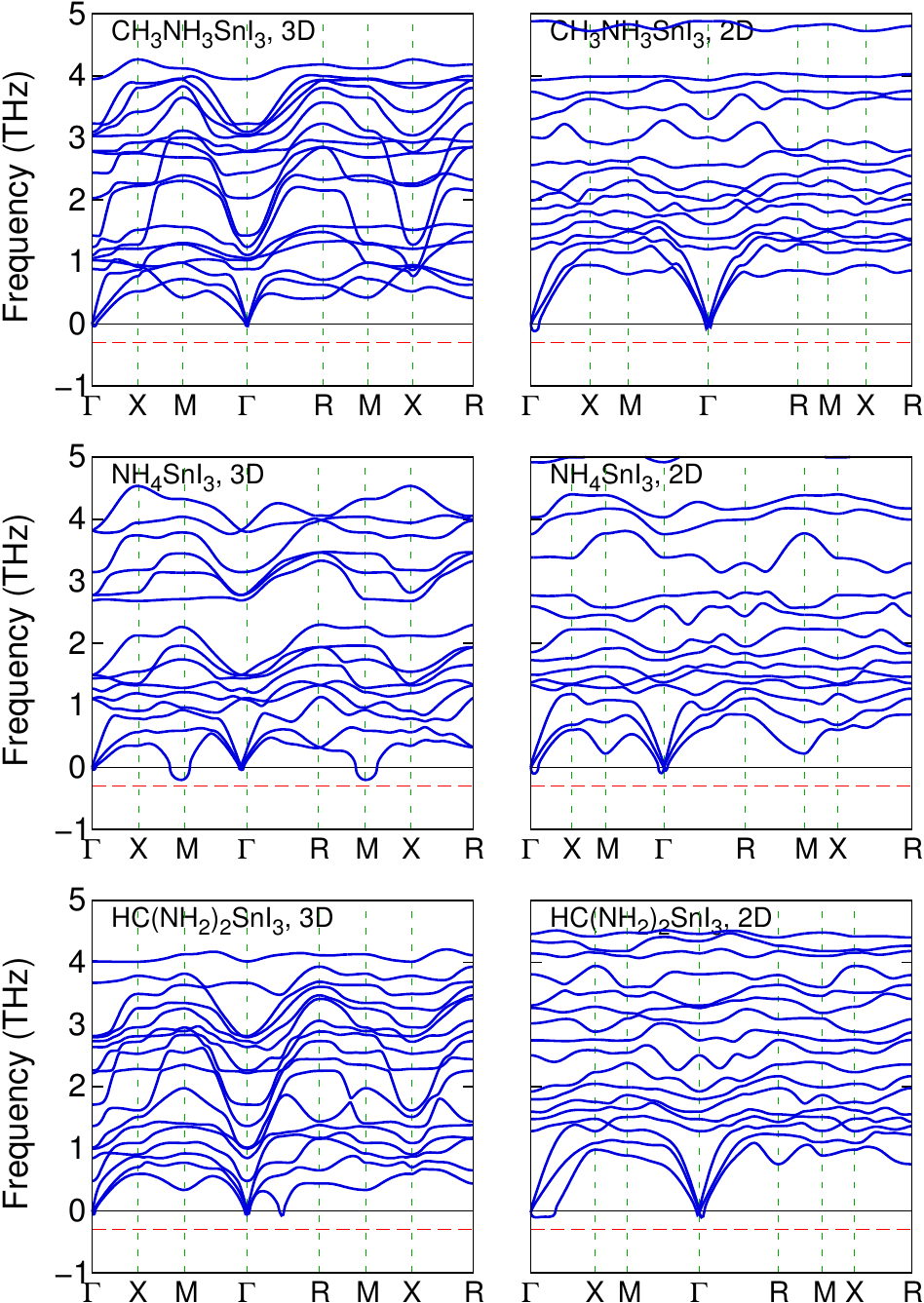}
  \caption{(Color online) Low-frequency regions of the calculated phonon band structures of the thermodynamically most stable 2D and 3D structures predicted for CH$_3$NH$_3$SnI$_3$, NH$_4$SnI$_3$, and HC(NH$_2$)$_2$SnI$_3$ (the full vibrational frequency spectra of these structures span up to $\sim 100$ THz). For convenience, phonon bands with imaginary frequencies are shown as those with {\it negative} frequencies. Red dashed lines sketch the numerical error of $\simeq 0.3$ THz typically encountered by phonon calculations.} \label{fig:phband}
  \end{center}
\end{figure}

For any hybrid perovskite, the lowest-energy 2D and 3D structures are energetically competing with rather small energy difference. In case of NH$_4$SnI$_3$, the 2D structure is lower than the 3D structure by $\simeq 6$ meV/atom. On the other hand, the 2D structure of CH$_3$NH$_3$SnI$_3$ and HC(NH$_2$)$_2$SnI$_3$ is higher than the 3D structures by roughly $7$ meV/atom and $1$ meV/atom, respectively. To further confirm the thermodynamic stability of the 2D structures with respect to the 3D structures, we performed additional calculations for the DFT energy $E_{\rm DFT}$ using different numerical schemes, employing the LDA, the Perdew-Burke-Ernzerhof (PBE),\cite{PBE} and the hybrid Heyd-Scuseria-Ernzerhof (HSE06)\cite{HSE,HSE06:2} XC functionals (without SOC). To somehow capture the long-range dispersion interaction, we also used a prescription which incorporates the vdW-DF2 non-local density functional.\cite{vdW-DF2} The results are shown in Table \ref{table:energy}, indicating that the 2D and 3D motifs of HC(NH$_2$)$_2$SnI$_3$ are essentially the same in energy. For NH$_4$SnI$_3$, the 2D structure is slightly more stable than the 3D structure while for CH$_3$NH$_3$SnI$_3$, the 3D structure is slightly more stable than its 2D counterpart. For all three hybrid perovskites, the 2D and 3D structures are essentially comparable in energy.

The phonon frequency spectrum of a theoretically predicted structure is typically used for examining its dynamical stability. We used the supercell approach, as implemented in {\sc phonopy} package, \cite{phonopy_sc, phonopy} to perform the relevant frozen-phonon calculations for all the six structures structures examined. For each of them, a $2\times 2 \times 2$ supercell was used for the phonon calculations. We note that calculations of this type are performed within the harmonic approximation on the structures which were determined at zero temperature. The low-frequency region (5 THz and  below) of the calculated phonon spectra is shown in Fig. \ref{fig:phband} while the full spectra (spanning up to $\simeq 100$ THz) are given in the Supplemental Material.\cite{supplement} Fig. \ref{fig:phband} indicates that within the known numerical error of phonon calculations ($\simeq 0.3$ THz, or roughly $0.1$ meV/atom for our structures),\cite{Huan:NaSc,VossMgBH4} the six structures indeed correspond to the local minima of the energy landscape, and hence, are dynamically stable.

\begin{figure}[t]
  \begin{center}
    \includegraphics[width= 7.5 cm]{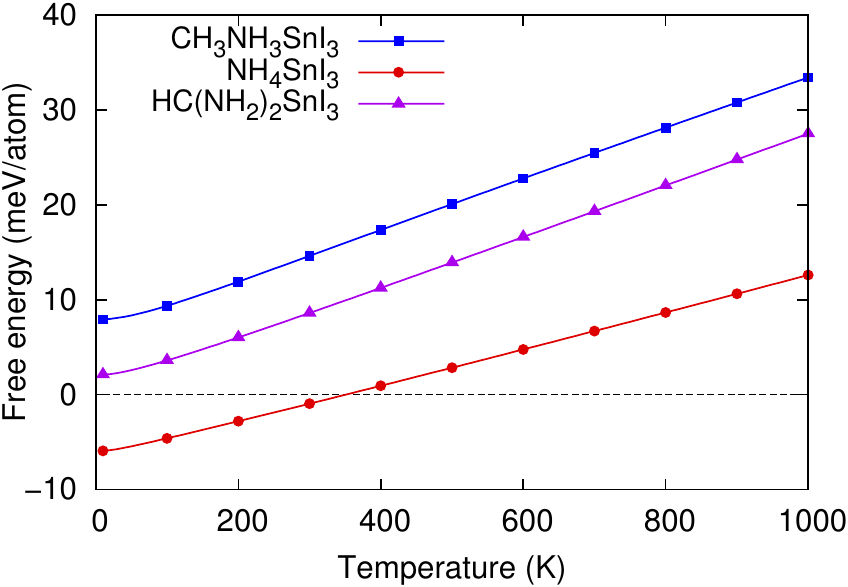}
  \caption{(Color online) Helmholtz free energy of the energetically most favorable 2D structures of CH$_3$NH$_3$SnI$_3$, NH$_4$SnI$_3$, and HC(NH$_2$)$_2$SnI$_3$, given with respect to that of the most stable 3D structure of the same material (dotted line). } \label{fig:free_energy}
  \end{center}
\end{figure}

At finite temperatures $T$, the thermodynamic stability of these structures may be accessed by estimating the Helmholtz free energy $F(T) = E_{\rm DFT}+F_{\rm vib}(T)$. \cite{Huan:Zn,Huan:Mixed,Huan:NaSc,Huan:hafnia} In this expression, $F_{\rm vib}(T)$ is the entropic contribution to $F(T)$ from the lattice vibrations. This term can be estimated by the lattice dynamics approach within the harmonic approximation from the calculated phonon density of states $g(\omega)$ as\cite{VossMgBH4,Ackland:freeenergy, Huan:freeenergy}
\begin{equation}\label{eq:fvib}
F_{\rm vib}(T) = 3Nk_{\rm B}T\int_0^\infty d\omega g(\omega)\ln\left[2\sinh\left(\frac{\hbar\omega}{2k_{\rm B}T}\right)\right].
\end{equation}
Here, $k_{\rm B}$ is the Boltzmann constant, $N$ the number of degrees of freedom, and $\omega$ the phonon frequency. The phonon density of state $g(\omega)$ used for Eq. \ref{eq:fvib} were evaluated by sampling the Brilluion zone by a $\bf q$-point mesh of $9\times 9\times 9$. We show in Fig. \ref{fig:free_energy} the free energy $F(T)$ calculated for the six structures examined, indicating that the 2D structure of NH$_4$SnI$_3$ is thermodynamically more stable than the 3D structure at low temperatures but gradually become less stable at $T\simeq 350$K and above. In cases of CH$_3$NH$_3$SnI$_3$ and HC(NH$_2$)$_2$SnI$_3$, the 3D structure is always favorable over the 2D structure within the temperature range examined. Overall, in all three materials, the energy differences between the 2D and 3D structures are relatively small and comparable with $k_{\rm B}T$, which is about 25 meV/atom at room temperature.
\begin{figure}[t]
  \begin{center}
    \includegraphics[width= 8 cm]{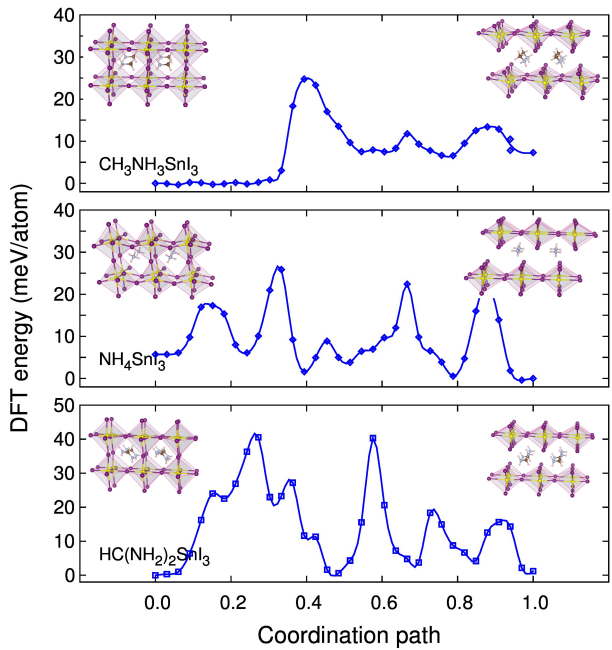}
  \caption{(Color online) Computed minimum energy pathways from the 3D (left) to the 2D (right) structures of CH$_3$NH$_3$SnI$_3$, NH$_4$SnI$_3$, and HC(NH$_2$)$_2$SnI$_3$. Curves were interpolated from the calculated energies of the images, given by symbols.} \label{fig:neb}
  \end{center}
\end{figure}

\section{Activation barriers}
Because the 2D and 3D motifs of CH$_3$NH$_3$SnI$_3$, NH$_4$SnI$_3$, and HC(NH$_2$)$_2$SnI$_3$ are nearly equal in energy, it is interesting to examine the energy barriers between them. To find this information, we employed the solid-state (climbing image) nudged elastic band method \cite{CINEB,SSNEB} for estimating the minimum energy pathways between the 3D and the 2D structures of the hybrid perovskites. The calculated results are shown in Fig. \ref{fig:neb}, demonstrating that for these materials, the 3D and 2D motifs are separated by low energy barriers, falling between $30-40$ meV/atom. The structural transition from a 3D to a 2D structure involves a series of steps, including breaking the long Sn-I ``bonds" along the out-of-plane direction, rotating both the in-plane Sn-I bonds and the whole organic cations, and then forming new Sn-I bonds to construct the 2D motif.\cite{supplement} Moreover, all of the organic cations are rigid and highly anisostropic, so their rotations will not be smooth. Consequently, the transformation from a 3D to a 2D structure generally encounters a sequence of barriers, as can be seen in Fig. \ref{fig:neb}. At operating temperatures (room temperature and above), these barriers are however small, being comparable with $k_{\rm B}T$.

\section{Electronic structures}\label{sec:elec}

\begin{figure}[t]
  \begin{center}
    \includegraphics[width= 7.75 cm]{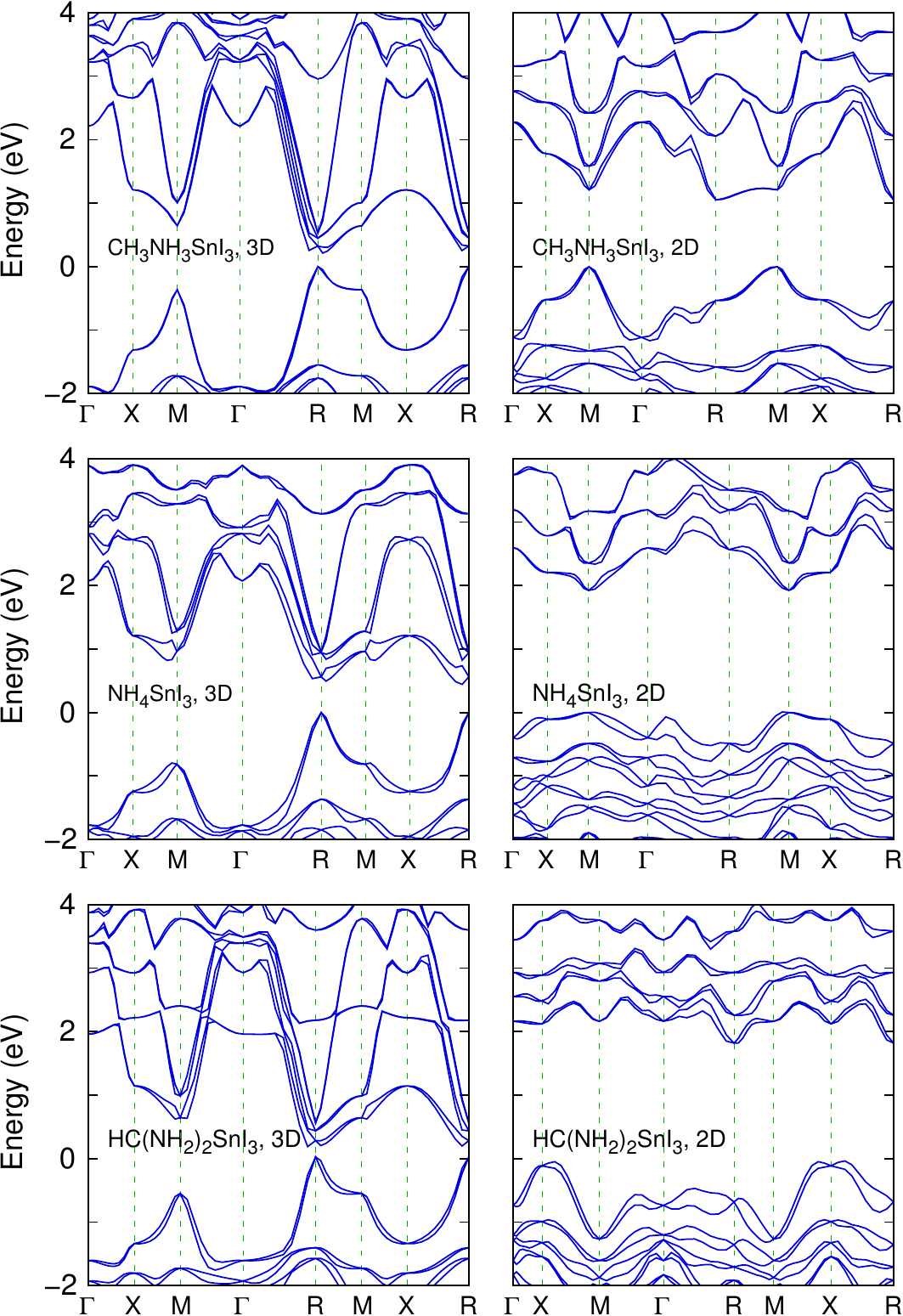}
  \caption{(Color online) Computed electronic bands of the 3D and 2D structures of CH$_3$NH$_3$SnI$_3$, NH$_4$SnI$_3$, and HC(NH$_2$)$_2$SnI$_3$. Fermi energies are set to zero.} \label{fig:band}
  \end{center}
\end{figure}

We performed electronic structure calculations for the 2D and 3D structures of the examined hybrid perovskites, and the results are shown in Fig. \ref{fig:band}. Interestingly, while the 3D structures are characterized by high-curvature parabolic bands, the 2D structures feature many flat bands  next to the conduction band minimum (CBM) and the valence band maximum (VBM) as previously addressed,\cite{Umebayashi:2003, Mosconi} thereby leading to some high-density electronic levels near the band edges. At the LDA+SOC level of DFT computations, the band gap $E_{\rm g}$ of the 2D structures is found (see Table \ref{table:bandgap}) to be considerably higher than that of the 3D counterparts while the CBM degeneracy is unambiguously lifted.

\begin{table}[t]
\caption{Calculated band gap of the 2D and 3D structures of three hybrid perovskites, given in eV. Results were obtained by using the HSE06 functional for the exchange-correlation energies while those obtained with LDA+SOC are given in parentheses for completeness.}\label{table:bandgap}
\begin{center}
\begin{tabular}{l c c c}
\hline
\hline
Material                   & 2D       & 3D      & Literature    \\
\hline
CH$_3$NH$_3$SnI$_3$        &  $3.00$ ($1.3$)   &  $1.22$ ($0.2$)  &  $1.20-1.35$\cite{Stoumpos_perovskite}    \\
NH$_4$SnI$_3$              &  $3.05$ ($1.9$)   &  $1.65$ ($0.4$) & $-$            \\
HC(NH$_2$)$_2$SnI$_3$      &  $2.75$ ($2.0$)   &  $1.20$ ($0.2$) & $1.41$\cite{Stoumpos_perovskite}          \\
\hline
\hline
\end{tabular}
\end{center}
\end{table}

As the intrinsic band gap underestimation of DFT with (semi)local XC functionals like LDA and PBE is inevitable,\cite{Perdew:bandgap} we performed additional calculations for the electronic structures with the hybrid HSE06 XC functional. The obtained electronic band structures are given in the Supplemental Material while the calculated band gaps are shown in Table \ref{table:bandgap}. For the 3D structure of CH$_3$NH$_3$SnI$_3$ and HC(NH$_2$)$_2$SnI$_3$, the calculated band gaps are $1.22$ eV and $1.20$ eV. These values are well consistent with the experimentally determined band gaps of 1.20 eV$-$1.35 eV and 1.41 eV, respectively.\cite{Stoumpos_perovskite} At the HSE06 level of DFT, the calculated band gap of the 2D structures is also significantly larger that that of the 3D structures, as revealed by calculations at the LDA+SOC level. Moreover, the electronic structures of the 3D structures of CH$_3$NH$_3$SnI$_3$ and HC(NH$_2$)$_2$SnI$_3$ (those experimentally realized) are rather similar, suggesting that HC(NH$_2$)$_2$SnI$_3$ may also be promising for the photovoltaic and thermoelectric applications of CH$_3$NH$_3$SnI$_3$.

\section{Summary}
In summary, we have systematically studied three organic-inorganic hybrid halide perovskites, i.e., CH$_3$NH$_3$SnI$_3$, NH$_4$SnI$_3$, and HC(NH$_2$)$_2$SnI$_3$, by first-principles calculations. We find that in addition to the 3D network of BX$_6$ octahedra, the 2D motif formed by alternating layers of BX$_6$ octahedra and organic cations may also be favorable. When the ionic cations are large, the layered structures would become dominant, as unambiguously observed in experiments.\cite{Mitzi, Mitzi95, Xu:layer, Kagan99, Xu:03} While these two structural motifs are energetically competing in CH$_3$NH$_3$SnI$_3$, NH$_4$SnI$_3$, and HC(NH$_2$)$_2$SnI$_3$, they are separated by low barriers ($\simeq 30-40$ meV/atom). The 2D layered structures are significantly different from the conventional 3D structures in terms of the electronic structures. On the other hand, the electronic structures of the 3D motifs of CH$_3$NH$_3$SnI$_3$ and HC(NH$_2$)$_2$SnI$_3$ are quite similar. In regard of the photovoltaic and thermoelectric applications of the hybrid halide perovskites, the present results suggest that further in-depth investigations of formamidnium tin iodide HC(NH$_2$)$_2$SnI$_3$ and the layered structures of the hybrid perovskites are desirable.

\begin{acknowledgements}
The authors thank Stefan Goedecker and Max Amsler for the minima-hopping code, and Rampi Ramprasad, Michael Pettes and Lilia Woods for useful discussions. They also thank two anonymous reviewers, whose comments have led to some significant improvements of the paper. Part of this work (by V.N.T.) is supported by the Vietnam National Foundation for Science and Technology Development under Grant 103.01-2014.25. Computational work by T.D.H. was made possible through the XSEDE computational resource allocation number TG-DMR150033.
\end{acknowledgements}


\end{document}